\documentclass[onecolumn]{aastex631}

\graphicspath{{./}{figures/}}

\begin{document}

\title{First evidence of multi-iron sub-populations in the Bulge Fossil Fragment candidate Liller 1\footnote{Based on observations collected at the Very Large Telescope of the European
  Southern Observatory, Cerro Paranal (Chile), under the
  ESO-VLT Multi-Instrument Kinematic Survey (MIKiS survey) 
programmes 106.21N5 and 105.20B9 (PI: Ferraro), and under 
MUSE science verification programmes: 60.A-9489;60.A-9343.}}

\author[0009-0002-8571-5170]{Chiara Crociati}
\affiliation{Dipartimento di Fisica e Astronomia, Universit\`a di Bologna, Via Gobetti 93/2 I-40129 Bologna, Italy}
\affiliation{INAF-Osservatorio di Astrofisica e Scienze dello Spazio di Bologna, Via Gobetti 93/3 I-40129 Bologna, Italy}

\author[0000-0002-6092-7145]{Elena Valenti}
\affil{European Southern Observatory, Karl-Schwarzschild-Strasse 2, 85748 Garching bei M\"{u}nchen, Germany}
\affil{Excellence Cluster ORIGINS, Boltzmann-Strasse 2, D-85748 Garching bei M\"{u}nchen, Germany}

\author[0000-0002-2165-8528]{Francesco R. Ferraro}
\affil{Dipartimento di Fisica e Astronomia, Universit\`a di Bologna, Via Gobetti 93/2 I-40129 Bologna, Italy}
\affil{INAF-Osservatorio di Astrofisica e Scienze dello Spazio di Bologna, Via Gobetti 93/3 I-40129 Bologna, Italy}

\author[0000-0002-7104-2107]{Cristina Pallanca}
\affil{Dipartimento di Fisica e Astronomia, Universit\`a di Bologna, Via Gobetti 93/2 I-40129 Bologna, Italy}
\affil{INAF-Osservatorio di Astrofisica e Scienze dello Spazio di Bologna, Via Gobetti 93/3 I-40129 Bologna, Italy}

\author[0000-0001-5613-4938]{Barbara Lanzoni}
\affil{Dipartimento di Fisica e Astronomia, Universit\`a di Bologna, Via Gobetti 93/2 I-40129 Bologna, Italy}
\affil{INAF-Osservatorio di Astrofisica e Scienze dello Spazio di Bologna, Via Gobetti 93/3 I-40129 Bologna, Italy}

\author[0000-0002-5038-3914]{Mario Cadelano}
\affil{Dipartimento di Fisica e Astronomia, Universit\`a di Bologna, Via Gobetti 93/2 I-40129 Bologna, Italy}
\affil{INAF-Osservatorio di Astrofisica e Scienze dello Spazio di Bologna, Via Gobetti 93/3 I-40129 Bologna, Italy}

\author[0000-0002-4639-1364]{Cristiano Fanelli}
\affil{INAF-Osservatorio di Astrofisica e Scienze dello Spazio di Bologna, Via Gobetti 93/3 I-40129 Bologna, Italy}

\author[0000-0002-6040-5849]{Livia Origlia}
\affil{INAF-Osservatorio di Astrofisica e Scienze dello Spazio di Bologna, Via Gobetti 93/3 I-40129 Bologna, Italy}
 
\author[0000-0001-9545-5291]{Silvia Leanza}
\affil{Dipartimento di Fisica e Astronomia, Universit\`a di Bologna, Via Gobetti 93/2 I-40129 Bologna, Italy}
\affil{INAF-Osservatorio di Astrofisica e Scienze dello Spazio di Bologna, Via Gobetti 93/3 I-40129 Bologna, Italy}

\author[0000-0003-4237-4601]{Emanuele Dalessandro}
\affil{INAF-Osservatorio di Astrofisica e Scienze dello Spazio di Bologna, Via Gobetti 93/3 I-40129 Bologna, Italy}

\author[0000-0001-9158-8580]{Alessio Mucciarelli}
\affil{Dipartimento di Fisica e Astronomia, Universit\`a di Bologna, Via Gobetti 93/2 I-40129 Bologna, Italy}
\affil{INAF-Osservatorio di Astrofisica e Scienze dello Spazio di Bologna, Via Gobetti 93/3 I-40129 Bologna, Italy}

\author[0000-0003-0427-8387]{R.~Michael Rich}
\affiliation{Department of Physics and Astronomy, University of California, 90095 Los Angeles, CA, USA}

\begin{abstract}

In the context of a project aimed at characterizing the properties of the so-called Bulge Fossil
Fragments (the fossil remnants of the bulge formation epoch), here we present the first determination
of the metallicity distribution of Liller 1. For a sample of 64 individual member stars we used ESO-
MUSE spectra to measure the equivalent width of the CaII triplet and then derive the iron abundance.
To test the validity of the adopted calibration in the metal-rich regime, the procedure
was first applied to three reference bulge globular clusters (NGC 6569, NGC 6440, and NGC
6528). In all the three cases, we found single-component iron distributions, with abundance values fully
in agreement with those reported in the literature. The application
of the same methodology to Liller 1 yielded, instead, a clear bimodal
iron distribution, with a sub-solar component at $\text{[Fe/H]}= -0.48\,$dex ($\sigma =
0.22$) and a super-solar component at $\text{[Fe/H]}= +0.26\,$dex ($\sigma =
0.17$). The latter is found to be significantly more
centrally concentrated than the metal-poor population, as expected in a self-enrichment scenario and
in agreement with what found in another bulge system, Terzan 5. The obtained metallicity
distribution is astonishingly similar to that predicted by the reconstructed star formation history of
Liller 1, which is characterized by three main bursts and a low, but constant, activity of star formation
over the entire lifetime. These findings provide further support to the possibility that, similar to Terzan
5, also Liller 1 is a Bulge Fossil Fragment.

\end{abstract}


\section{Introduction}
\label{sec:intro}
The formation of galaxy bulges is currently highly debated in the
literature: several different mechanisms, ranging from dissipative
collapse, to dynamical secular evolution of unstable discs and
merging/coalescence of primordial substructures, have been proposed
\citep[e.g.,][]{combes+1990, immeli+2004, elmegreen+2009}.  
On the other hand, the detection of the
so-called ``clumpy or chain galaxies'' observed at high-redshift
\citep[e.g.,][]{carollo+2007, elmegreen+2009} suggests that the
coalescence of primordial substructures is a promising channel that
could have played a relevant role in the assembling process of
galactic spheroids. Indeed, numerical simulations
\citep[e.g.,][]{immeli+2004, elmegreen+2008, bournaud+2009,
  bournaud2016} have shown that primordial massive clumps (with
masses of $10^{8-9} M_\odot$) can form from violent disk instabilities
and eventually migrate to the centre and dissipatively coalesce
generating the bulge.  The same simulations also show that, while the
majority of such primordial clumps coalesces to form the bulge, a few
of them can survive the total disruption and be still present in the
inner regions of the host galaxy, roughly appearing as massive
globular clusters (GCs). At odds with genuine GCs, however, these
fossil relics are expected to host multi-iron and multi-age
sub-populations, because their progenitors were massive enough to
retain the iron-enriched ejecta of supernova (SN) explosions and
likely experienced multiple bursts of star formation. This
scenario should hold also for the formation of the Milky Way spheroid
and, intriguingly, two peculiar stellar systems promising to be the
fossil records of this hierarchical assembly process have been
recently discovered in the Galactic bulge.

The first candidate, hidden under the false identity of a massive
($2\times 10^6 M_\odot$; \citealt{lanzoni+10}) GC named Terzan 5, was
identified back in 2009. Its detailed photometric and spectroscopic
study \citep{ferraro+09,massari+14,origlia+11,origlia+13,origlia+19}
demonstrated that it hosts at least two major sub-populations,
ascribable to different star formation events \citep{ferraro+16}: the
first occurred $\sim 12$ Gyr ago, at the epoch of the Galaxy
assembling, while the second is much more recent (dating back to $\sim
4.5$ Gyr ago). A third, minor (possibly older) component with
$\text{[Fe/H]}=-0.79\,$dex and [$\alpha$/Fe]$= +0.36\,$dex seems also to be present
\citep{origlia+13}. Spectroscopic investigations have clearly shown
that two sub-solar populations formed out of gas exclusively enriched
by type II SNe (SNeII) up to $\text{[Fe/H]}\sim -0.3\,$dex, which is typical of
massive and dense environments that experienced star formation at very
high rates (as galaxy bulges and the high-redshift massive clumps
mentioned above). The youngest population is more centrally segregated
and has super-solar metallicity ($\text{[Fe/H]}=+0.3\,$dex) and solar-scaled
[$\alpha$/Fe], suggesting that the progenitor system (the proto-Terzan
5) was massive enough (as the high-redshift clumps) to retain gas
ejected by both SNeII and SNeIa, before igniting a second burst of
star formation. The [$\alpha$/Fe] \emph{vs} [Fe/H] pattern drawn by
the sub-populations of Terzan 5 is perfectly consistent with that of
bulge field stars, while it is incompatible with those of the Milky
Way halo and Local Group dwarf galaxies. Indeed, the observed pattern
unambiguously demonstrates the Terzan 5 kinship with the bulge and
strongly supports an in-situ origin, thus classifying it as a valuable
candidate remnant of a massive clump that contributed to generate our
spheroid.  This scenario was further supported by the discovery
\citep{ferraro+21} that another GC-like object in the bulge (Liller 1)
hosts two distinct populations with remarkably different ages: 12 Gyr
for the oldest one, just 1-2 Gyr for the youngest component.  The
spectroscopic information currently available for Liller 1 suggests
that its old population has a chemistry fully compatible with that
measured for the old population of Terzan 5: $\text{[Fe/H]}= -0.3\,$dex and
[$\alpha$/Fe]$=+0.3\,$dex \citep{origlia+02}. However, this is based on
just two giant stars, and no information is available for the young
component, although photometric evidence suggests that it could be
super-solar \citep{ferraro+21, dalessandro+22}. In fact the
reconstructed star formation history of Liller 1 recently derived from
the analysis of its color-magnitude diagram (CMD) suggests the
occurrence of three main bursts producing an overall bimodal iron
abundance distribution \citep{dalessandro+22}. Clearly, the detailed
chemical characterization (in terms of iron and $\alpha-$elements
abundances) of the stellar populations in Liller 1 is urgent and of
paramount importance: firmly assessing that Liller 1 is a multi-iron
stellar system with a tight chemical connection to the bulge would
strongly indicate that, like Terzan 5, it is a Bulge Fossil Fragment,
i.e. the living remnant of one of the primordial massive clumps that
12 Gyr ago contributed to the Galactic bulge formation.

As first step of this investigation, here we take advantage of the
performance of the Multi Unit Spectroscopic Explorer (MUSE) at the
Very Large Telescope (VLT) of the European Southern Observatory (ESO)
to perform a preliminary screening of the iron abundance in a large
number of stars in Liller 1. The iron abundance has been estimated
from the equivalent width (EW) of CaII triplet (CaT) lines and
adopting calibration relations provided in the literature.
%% visible in the infrared wavelength range has been found to be a
%% powerful indicator of a star metallicity and many studies provided
%% calibration relations for deriving [Fe/H] values from CaT lines
%% equivalent widths (EWs).
Specifically, \citet{Husser2020} provide a CaT-metallicity relation
expressly calibrated for MUSE spectra, which links the EW of CaT
lines at 8542 and 8662 \AA\ and the difference between the star
magnitude and the horizontal branch (HB) level in the HST F606W
filter, with the iron abundance [Fe/H]. The relation has been
calibrated using 19 Galactic GCs (from \citealt{Dias2016}) with
metallicities ranging from $\text{[Fe/H]}\sim-2.3\,$dex up to
$\text{[Fe/H]}\sim-0.4\,$dex. However, photometric \citep{ferraro+21} and
spectroscopic \citep{origlia+02,origlia+97} studies suggest that the stellar
populations in Liller 1 could be more metal rich than
$\text{[Fe/H]}\sim-0.4\,$dex.  Therefore, as sanity check, we fist tested the
validity of this relation by analysing the MUSE spectra of three
metal-rich bulge clusters with [Fe/H] measurements from
high-resolution spectroscopic studies, namely NGC 6569
($\text{[Fe/H]}\sim-0.8\,$dex; \citealt{Valenti6624_6569}), NGC 6440 
($\text{[Fe/H]}\sim-0.5\,$dex; \citealt{Origlia6440}), and 
NGC 6528 ($\text{[Fe/H]}\sim-0.17\,$dex; \citealt{Origlia6528}). Then, we
applied the method to Liller 1.  The paper is organized as follows. In
Section~\ref{sec:dataset} we present the observational dataset. In Section~\ref{sec:data_analysis} we
describe the procedures adopted for the data reduction and the
relation adopted to derive the [Fe/H] abundances.  The results are
presented in Section~\ref{sec:results}, while Section~\ref{sec:summary_conclusions} is devoted to the discussion
and conclusions.

%----------------------------------------
\section{Dataset}
\label{sec:dataset}
%In this study we analyzed data acquired with the panoramic
%integral-field spectrograph MUSE mounted at ESO-VLT.  This instrument
%can be used in the Wide Field Mode (WFM), with a field of view of
%approximately $1\arcmin\times 1\arcmin$ and spatial sampling of
%$0.2\arcsec$/pixel, or in Narrow Field Mode (NFM), with a field of
%view of $7.5\arcsec\times7.5\arcsec$ and a resolution of
%$0.025\arcsec$/pixel.  The spectrograph is made of a mosaic of 24
%Integral Field Units (IFUs) and is also provided with an adaptive
%optic system.  An image slicer in front of each IFU serves as entrance
%slit, thus producing a spatially resolved spectrum.  The spectra cover
%the wavelength range 4800-9300 \AA, have a resolving power $R\sim
%3000$ at $\lambda\sim 8000$ \AA, and a spectral sampling of 1.25
%\AA.
In this study we analized data acquired with the integral field spectrograph MUSE mounted ad the ESO-VLT. The instrument is composed of 24 identical IFU modules that - when using the Wide Field Mode (WFM) - cover a field of view of $1\arcmin\times 1\arcmin$ , with a spatial sampling of $0.2\arcsec$/pixel. WFM observations can be performed either with natural seeing (i.e., WFM-noAO), or combined with the Ground Layer Adaoptive Optics mode (i.e., WFM-AO) of the VLT Adaoptive Optics Facility (AOF, \citealt{arsenault+2008}) through the GALACSI AO module \citep{stroebele+2012}.
In addition, GALACSI enables the so-called Narrow Field Mode (NFM): a $7.5\arcsec\times7.5\arcsec$ Laser Tomography AO corrected field of view sampled at $25\,$mas/pixel.
Spectrally, MUSE covers most of the optical range - i.e., 4800 - 9300 \AA\ (nominal filter) and 4650 - 9300 \AA\ (extended filter) - with a sampling of 1.25 \AA\ and a resolution of $R\sim3100$ at 8000 \AA.

In the following we report the details concerning the datasets
analyzed for the three reference GCs and Liller 1:
\begin{itemize}
\item NGC 6528 - For this cluster, we used four WFM-noAO archival
  observations that are part of the WFM science verification run
  (program ID: 60.A-9343(A)). Each exposure is 150 s long, and was
  secured with a DIMM seeing of $\sim 0.85\arcsec$. The secured MUSE
  pointing was roughly centered on the center of the cluster.
\item NGC 6440 - The dataset for this cluster has been secured as
  part of the Multi Instrument Kinematic Survey of Galactic GCs
  (ESO-MIKiS survey; see \citealt{ferraro+18a, ferraro+18b,
    lanzoni+18a, lanzoni+18b, leanza+23}), a spectroscopic survey
  aimed at using the current generation of spectrographs mounted at
  the VLT to characterize the internal kinematics of a representative
  sample of GCs.  Specifically, the data were acquired during the NFM
  science verification run (program ID: 60.A-9489(A), PI: Ferraro) and
  they consist of a mosaic of four MUSE/NFM pointings sampling
  approximately the innermost $15\arcsec$ of the cluster. This dataset
  has been presented in \cite{leanza+23} to discuss the kinematic
  properties of NGC 6440.  Here we analyzed the northern pointing with
  respect to the cluster center. Each exposure is  850 s long and the DIMM
seeing during the observations ranged from $0.45\arcsec$ to $0.8\arcsec$.
\item NGC 6569 - The observations of NGC 6569 are also part of
  the ESO-MIKiS survey and have been acquired under the Large Program
  ID: 106.21N5.003 (PI: Ferraro). They consist of seven MUSE/NFM
  pointings suitably displaced to sample the innermost $10\arcsec$
  from the cluster center. For each pointing, three $750$ s long
  exposures were obtained, with a resulting DIMM seeing better than
  $0.7\arcsec$. In this case all the 7 pointings were used. The detailed description 
of this dataset and the kinematic analysis of this cluster can be found in \citet{pallanca+23}.
\item Liller 1 - The dataset was acquired within the program  
  105.20B9 (PI: Ferraro) and is composed of five exposures in the
  MUSE/WFM configuration centered on the cluster center.  All the
  exposures are 880 s long, with an average DIMM seeing ranging from
  $0.8\arcsec$ to $0.9\arcsec$. 
\end{itemize}
 
\begin{figure}[t]
    \centering
    \includegraphics[width=1\textwidth]{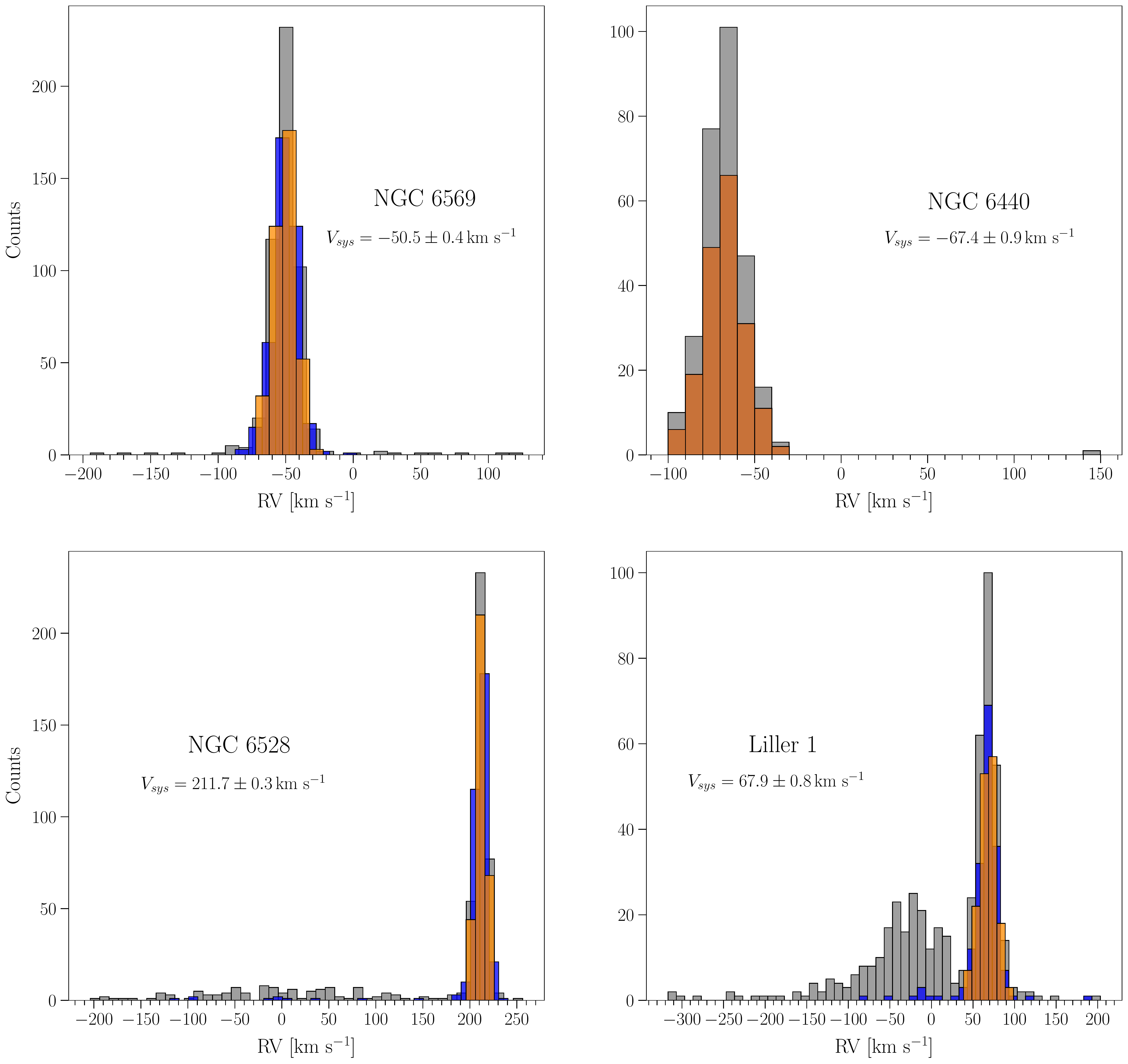}
    \caption{The grey histrograms show the RV distributions for all
      the stars extracted with PampelMuse (i.e., having spectra with
      $S/N \geq 10$) in the three reference clusters (NGC 6569, NGC
      6440 and NGC 6528) and in Liller 1 (from top-left, to
      bottom-right: see labels). The histograms shown in blue and
      orange colors correspond, respectively, to the sub-sample of
      PM-selected member stars, and the one further refined with a $3
      \sigma$-rejection applied to the measured RVs.  The mean RV ($V_{sys}$),
      with the relative error, of the orange distributions are also labelled in the panels.}
    \label{fig: RVs}
\end{figure}

%----------------------------------------
\section{Data Analysis}
\label{sec:data_analysis}
For all the stellar clusters analyzed in this study, we adopted the
same data analysis procedure, which can be summarized in five main
steps: (1) reduction of the MUSE data and combination of the multiple
exposures into the final datacube, (2) extraction of the spectra from
the MUSE datacube,
%by means of the software PampelMuse  \citep{KamannPampelMuse}; 
(3) measure of the radial velocity (RV) of each of the extracted spectra,
%following the procedure presented in \cite{leanza+23}; 
(4) measure of the EW of the two strongest CaT lines, and (5)
determination of the star metallicity.
%using the CaT-metallicity relation presented in \cite{Husser2020}.

{\it (1) Reduction of the MUSE data} - The data reduction was
performed making use of the most recent version of the standard MUSE
pipeline \citep{MUSEpipeline}. In this step, bias subtraction, flat
fielding, wavelength calibration, sky subtraction, astrometric and
flux calibration, and heliocentric velocity correction are performed
for each exposure of each individual IFU. Subsequently, the processed
data coming from all the 24 IFUs are combined in a single
datacube. Finally, we combined all the available exposures together,
obtaining the final datacube for each stellar cluster.

\begin{figure}[t]
    \centering
    \includegraphics[width=1\textwidth]{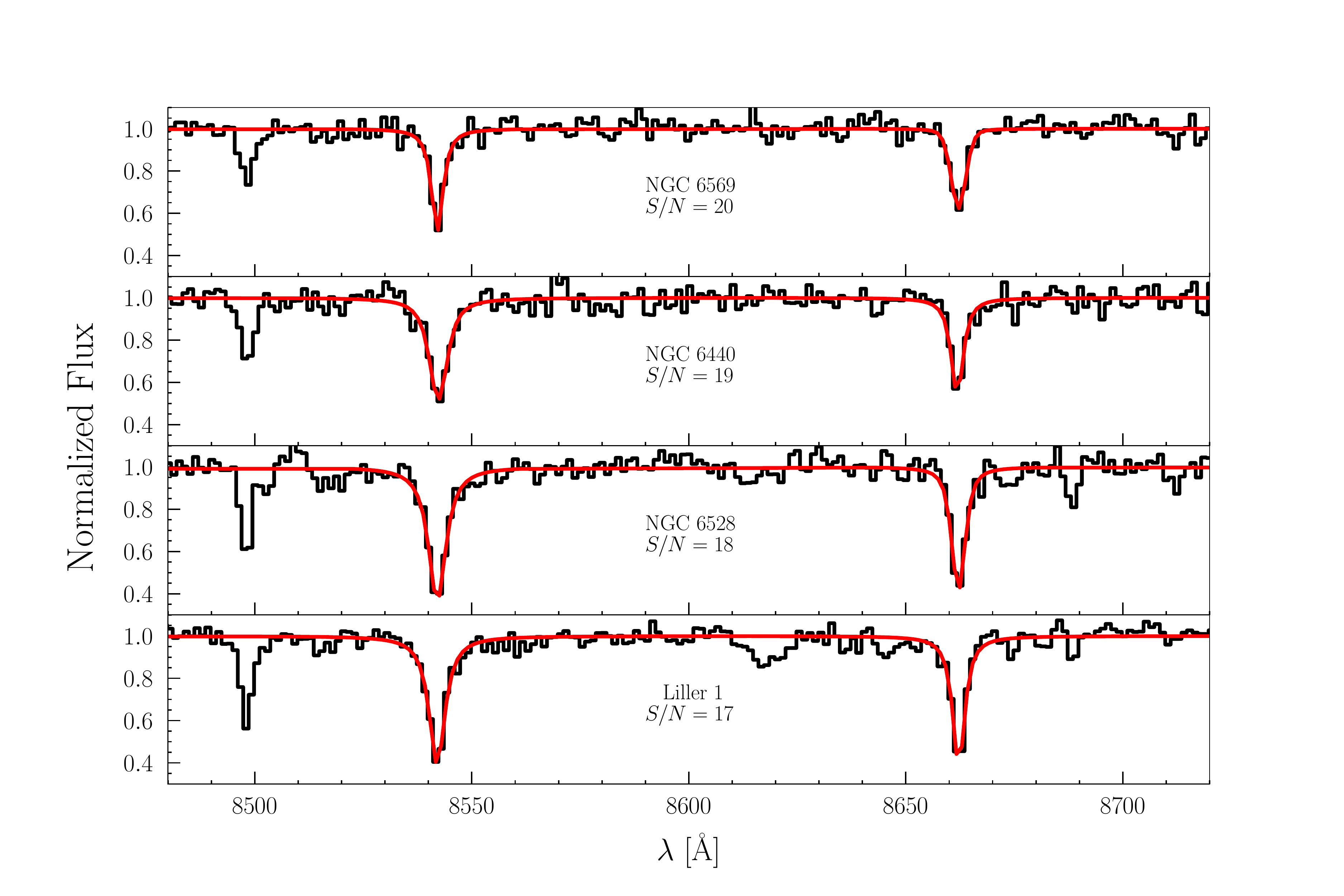}
    \caption{Examples of normalized spectra with $S/N > 15$ for each
      of the 4 stellar systems analyzed in this work. The observed
      spectra are shown in black, while the best fitting Voigt models 
for the two strongest CaT lines are overplotted in red.}
    \label{fig: fit_spectra}
\end{figure}

\begin{figure}[t]
    \centering
    \includegraphics[width=0.8\textwidth]{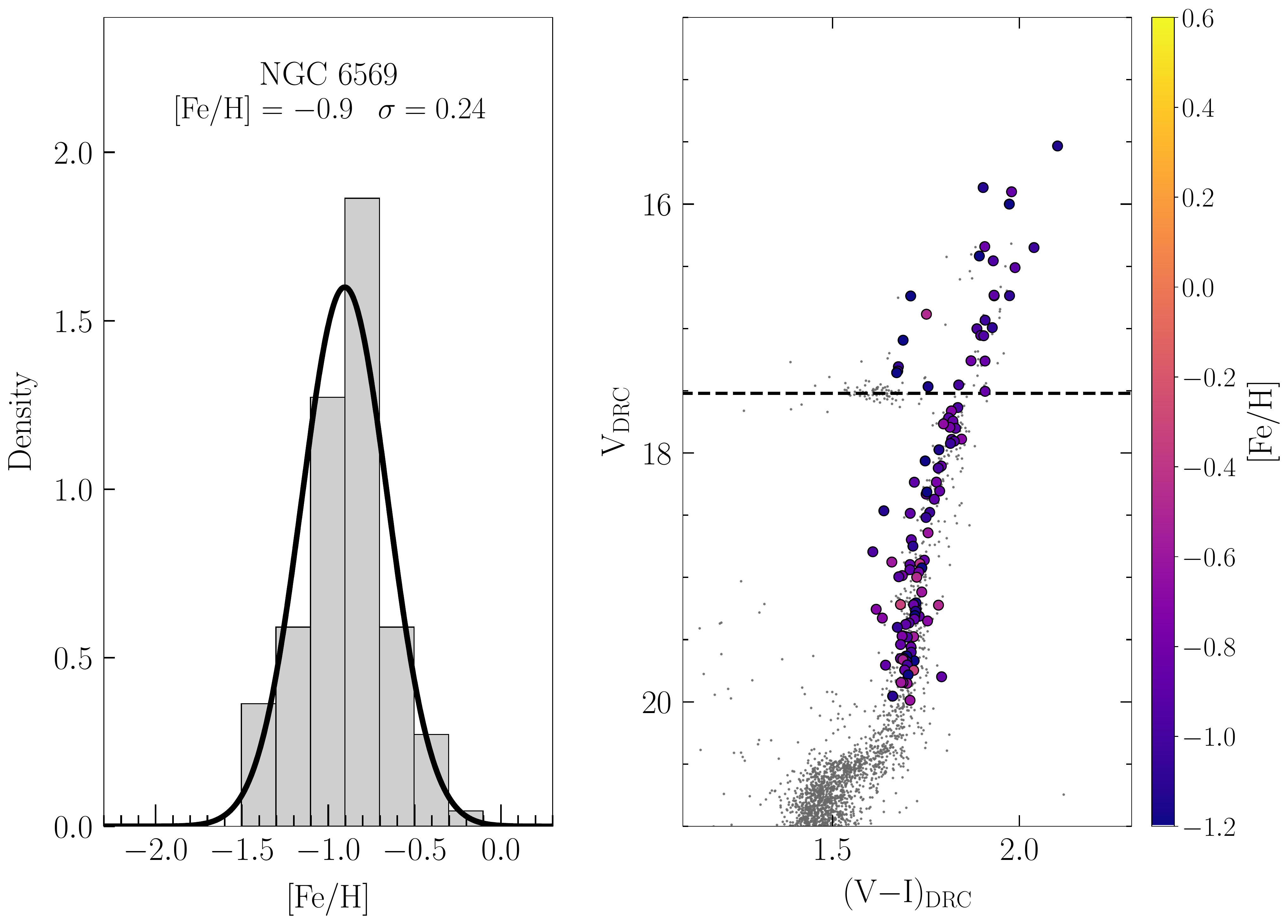}
    \caption{\emph{Left panel}: Metallicity distribution of NGC 6569
      (grey histogram) and its best-fit Gaussian solution (black
      line). The distribution is normalized such that the total area 
of the histogram equals $1$. The mean [Fe/H] value and the standard deviation derived
      from the Gaussian fit are also labelled in the
      panel. \emph{Right panel}: PM selected and DRC CMD of NGC 6569
      (grey dots) from the photometric catalog of
      \cite{Saracino6569}. The stars for which the metallicity has
      been measured are plotted as large circles colored according to
      their [Fe/H] value (see the color bar on the right). The black
      dashed line marks the adopted magnitude level of the HB 
($V_{\rm HB} = 17.52$).}
    \label{fig: NGC6569_MDF}
\end{figure}

\begin{figure}[t]
    \centering
    \includegraphics[width=0.8\textwidth]{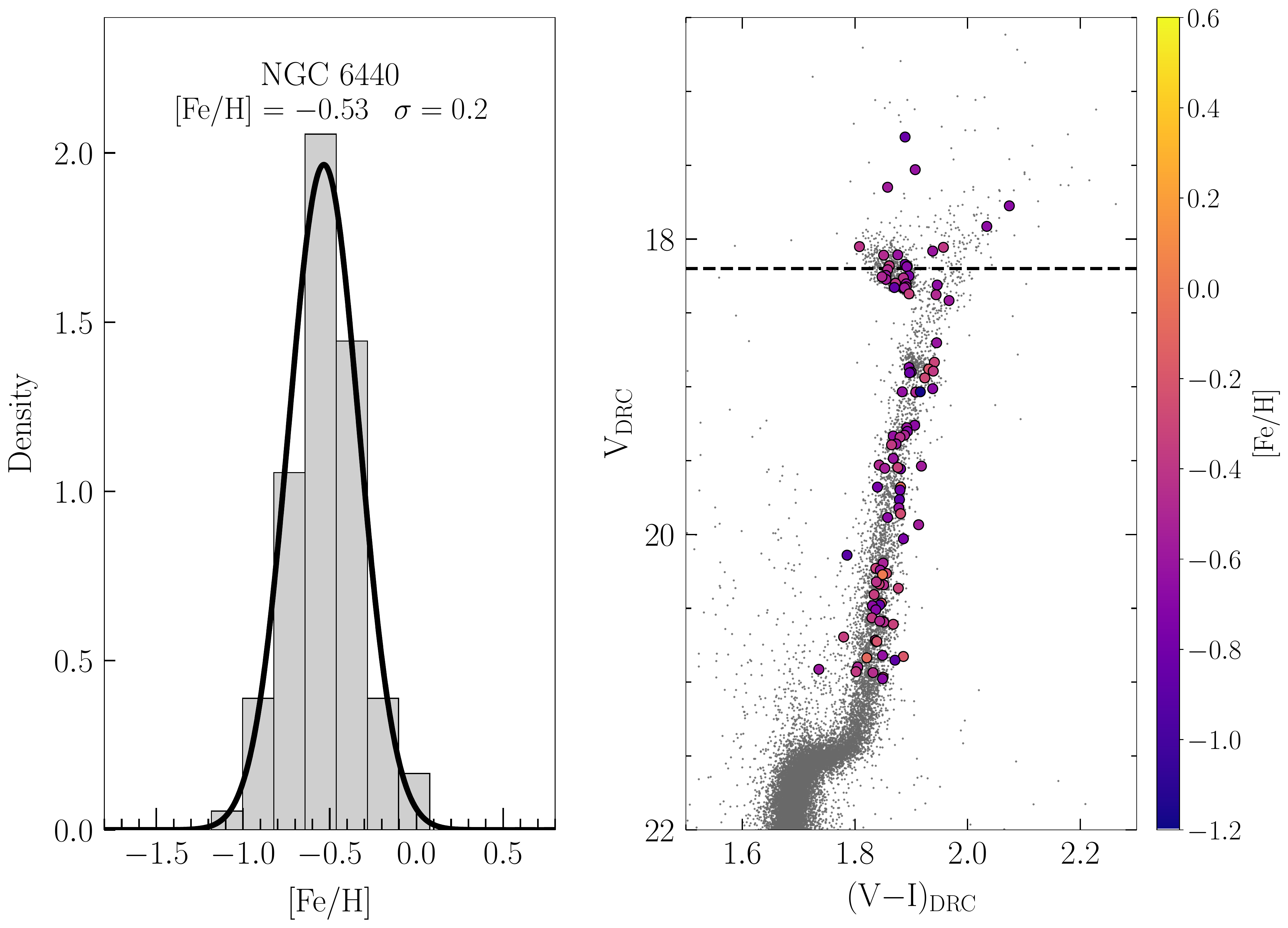}
    %% \caption{\emph{Left panel}: Metallicity distribution obtained from
    %%   MUSE spectra of NGC 6440 for the selected sample of stars as
    %%   described in Section~\ref{sec:validations_results}. The solid
    %%   line represents the fit that best reproduces the observed
    %%   distribution. The mean [Fe/H] value and the standard deviation
    %%   derived from the Gaussian fit is also reported in the
    %%   panel. \emph{Right panel}: PM- selected and DRC-CMD of NGC 6440
    %%   (grey points) from the photometric catalog of
    %%   \cite{Pallanca6440}. Stars considered for the metallicity
    %%   distribution are plotted as circles colored according to their
    %%   [Fe/H] value. The color code bar is shown on the right. The red
    %%   dashed line marks the HB level ($V_{HB} = 18.2$) that we assumed
    %%   to compute the magnitude difference of each star from the
    %%   HB. {\bf[B: fare NERI i punti nel CMD]}}
    \caption{As in Figure \ref{fig: NGC6569_MDF}, but 
for NGC 6440. Here we  considered the photometric catalog 
presented in \cite{Pallanca6440} and we assumed $V_{\rm HB} = 18.2$. }
    \label{fig: NGC6440_MDF}
\end{figure}

\begin{figure}[t]
    \centering
    \includegraphics[width=0.8\textwidth]{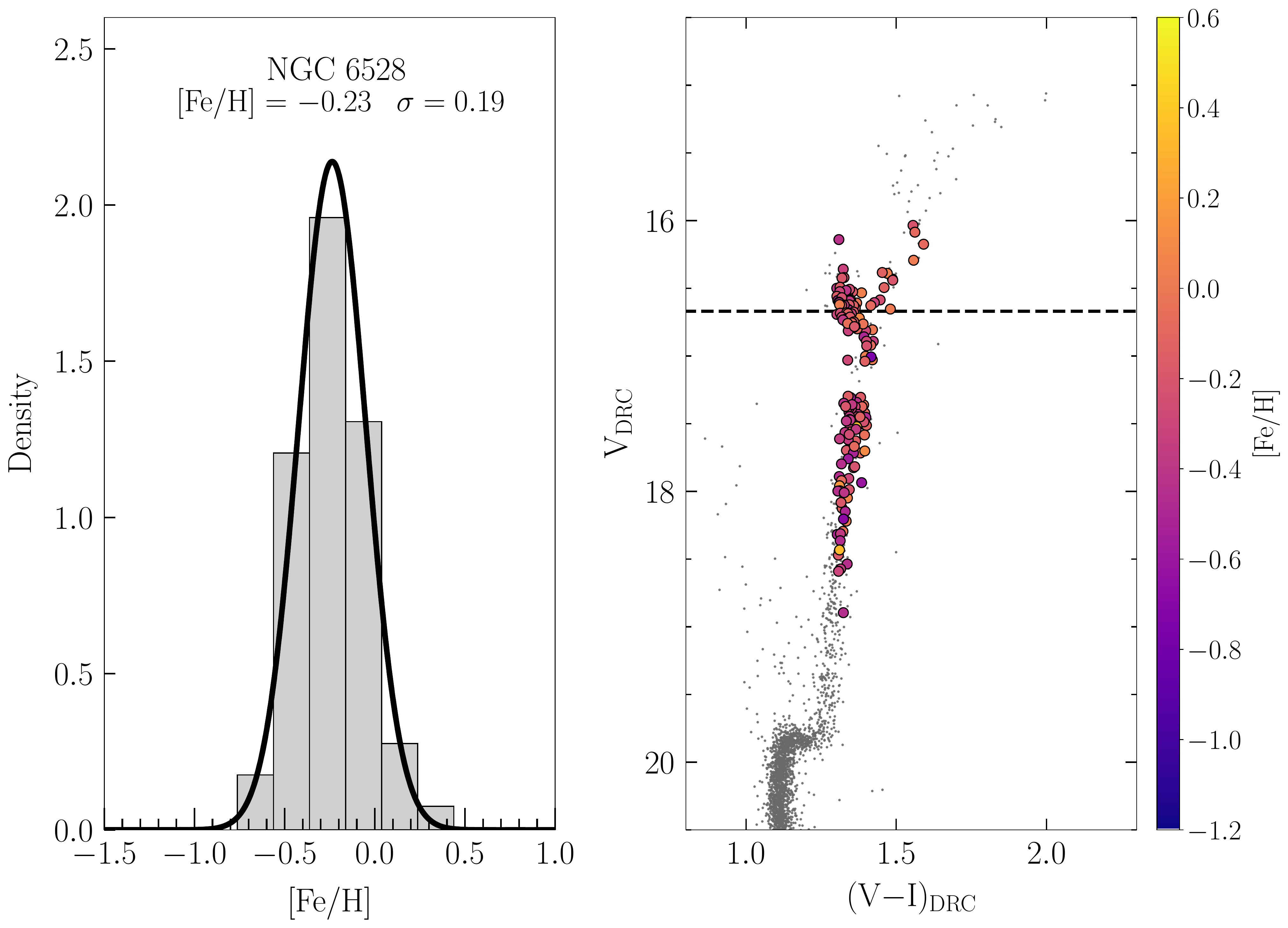}
    %% \caption{\emph{Left panel}: Metallicity distribution obtained from
    %%   MUSE spectra of NGC 6528 for the selected sample of stars as
    %%   described in Section~\ref{sec:validations_results}. The solid
    %%   line represents the fit that best reproduces the observed
    %%   distribution. The mean [Fe/H] value and the standard deviation
    %%   derived from the Gaussian fit is also reported in the
    %%   panel. \emph{Right panel}:PM- selected and DRC-CMD of NGC 6528
    %%   (grey points) the photometric catalog of \cite{LagioiaNGC6528}.
    %%   Stars for which the metallicity has been measured are plotted as
    %%   circles colored according to their [Fe/H] value. The color code
    %%   bar is shown on the right.  The red dashed line marks the HB
    %%   level ($V_{HB} = 16.7$) that we assumed to compute the magnitude
    %%   difference of each star from the HB. {\bf[B: fare NERI i punti
    %%       nel CMD]}}
    \caption{As in Figure \ref{fig: NGC6569_MDF}, but for NGC 6528.  Here we  considered the photometric catalog presented in \cite{LagioiaNGC6528} and we assumed $V_{\rm HB} = 16.7$. }
\label{fig: NGC6528_MDF}
\end{figure}

{\it (2) Extraction of the spectra from the MUSE datacube }- As second
step, we extracted individual star spectra from the final datacube by
using the software PampelMuse \citep{KamannPampelMuse}. This pipeline
is recommended for extracting spectra from observations of crowded
stellar fields, since it can perform source deblending through a
wavelength-dependent point spread function (PSF) fitting. For a
successful extraction, PampelMuse requires as input a fiducial
photometric catalog with high spatial resolution, high photometric
completeness and high astrometric accuracy. Hence, for each considered
cluster, we adopted an accurate photometric catalog obtained from HST
observations, also including differential reddening corrections and
proper motion (PM) information.
%% , and providing the required magnitude difference of each individual
%% star from the cluster HB level in the F606W filter.
In the case of NGC 6528, we used the photometric catalog published in
\citet{LagioiaNGC6528}, which includes F606W and F814W magnitudes
measured from the HST ACS/WFC camera. For NGC 6440, we adopted the
catalog presented in \citet{Pallanca6440}, based on HST/WFC3
observations in the same filters. This was also adopted in
\cite{leanza+23} for the extraction of the MUSE spectra for the
entire dataset discussed there, while here we used the spectra
acquired in the northern pointing only. The high resolution
photometric catalog available for NGC 6569 is the one presented in
\citet{Saracino6569}, providing HST/WFC3 optical images in the F555W
and F814W filters, together with near-infrared (NIR) $J$ and $K_s$
images acquired with the Gemini multi-conjugate adaptive optic system
GeMS. The photometric catalog used for Liller 1 is described in detail
in \citet{ferraro+21} and \citet{dalessandro+22}. It is based on
high-resolution HST ACS/WFC observations in the filters F606W and
F814W, once more combined with $J$ and $K_s$ images acquired with
Gemini/GeMS (see also \citealt{saracino+15}).
%% some of the brightest stars were missing and we thus recovered them
%% by {\bf ???? performing a photometric reduction on the HST dataset
%% of Liller 1 presented in \cite{ferraro+21}, applying their best PSF
%% model to the bright stars by using \texttt{DAOPHOTIV}
%% \citep{Stetson1987}.}

PampelMuse needs in input a specific photometric band and an
analytical PSF model. We choose as input magnitudes those in the F814W
filter, and as PSF model, the Moffat function in case of WFM
observations, and the MAOPPY function \citep{MaoppyFunction} in the
case of NFM data (see \citealt{leanza+23}
for additional details about the extraction and the reduction of the
NFM dataset). For the following analysis, we only considered extracted
spectra flagged as ``adequate'' by the software, which correspond to
a required signal-to-noise ratio $S/N\geq 10$.

{\it (3) measure of the radial velocity} - The next step is the determination 
of the stars RV, used to check and possibly further constrain the cluster 
membership already inferred from the available proper motions. For NGC 6440 and NGC
6569 we used the RV catalogs obtained, respectively, in
\citet{leanza+23} and \citet{pallanca+23}, while for 
Liller 1 and NGC 6528 we computed the RVs following
the procedure explained in \cite{leanza+23}. This is based on the
measure of the Doppler shift of the CaT lines from the comparison
between the observed stellar spectrum and the relative best-fit
synthetic model chosen from a library of templates. Our library was
composed of synthetic spectra generated with the SYNTHE code
(\citealt{Sbordone2004} and \citealt{Kurucz2005}) spanning a metallicity range from $-0.5\,$dex to
$+0.5\,$dex with a step of $0.25\,$dex, and temperatures varying from
$3750$ K to $4750$ K with a step of $250$ K. Once the observed
spectrum is normalized to the continuum by means of a spline fitting
in the 7300 - 9300 \AA\ wavelength range, the procedure computes the
residuals between the observed spectrum and each template shifted in
velocity in steps of 0.1 km s$^{-1}$. We considered as the best-fit
synthetic spectrum the one providing the distribution of the residuals
with the smallest standard deviation. As a consequence, the RV of the
star is obtained from the minimum of the distribution. As shown in \cite{ValentiRV} and \cite{leanza+23}, considering the metallicity range of our interest ($\text{[Fe/H]}\ge -1$), the typical uncertainty on the estimated of individual RV is $\sim8$ km s$^{-1}$ for the stars with the lowest $S/N$ ratio and it decreases up to $\sim1.5$ km s$^{-1}$ for the targets with the highest $S/N$.

In Figure \ref{fig: RVs} we show the RV distributions of all the spectra
extracted with PampelMuse (grey histograms). As can be seen, in the
considered cases the population of cluster members is clearly
distinguishable as a narrow, strongly peaked component, while the
bulge field component appears as a sparse population spanning a wide
range of RVs (see the case of NGC 6569), or define a broad
distribution peaking at a different mean value (see the cases of NGC
6528 and Liller 1).

The cluster systemic velocity ($V_{\rm sys}$) has been estimated as
the mean value of the sole likely cluster members after a
$3\sigma$-clipping rejection removing stars with clearly discrepant
RVs.  It is worth of emphasising that in the case of Liller 1 only 11
stars over the 171 PM-selected ones show RV inconsistent with the
cluster systemic velocity. This provide a further confirmation that
the PM-selected sample discussed in \citet{ferraro+21} and
\citet{dalessandro+22} is largely dominated by stars belonging to
Liller 1, and only marginally affected by residual bulge
contamination. By construction, for NGC 6569 and NGC 6440 we obtain
$V_{\rm sys}$ values (see labels) fully consistent with those quoted
in \cite{pallanca+23} and \cite{leanza+23}, respectively.  For NGC
6528 and Liller 1 we found $V_{\rm sys}=211.7\pm0.3$ km s$^{-1}$ and
$V_{\rm sys}=67.9\pm0.8$ km s$^{-1}$, respectively. While the value
obtained for NGC 6528 is in agreement with that ($V_{\rm
  sys}=211.86\pm0.43$ km s$^{-1}$) quoted in \citet{Baumgardt2018}, in
the case of Liller 1 it turns out to be significantly different
($V_{\rm sys}=60.36\pm2.44$ km s$^{-1}$), possibly due to a residual
contamination from field stars in the sample analysed by \citet{Baumgardt2018}.

The values of $V_{\rm sys}$ thus derived have been used for the
selection of member stars, taking into account the PM information and
requiring that the RV is within $3\sigma$ the cluster systemic
velocity.  Only for a few objects in Liller 1 for which no measured PM
was available, the membership is based only on the RV value. The
sample of member stars thus obtained counts a total of 387 stars in
NGC 6569, 184 stars in NGC 6440, 322 stars in NGC 6528, and 160 stars
in Liller 1.

{\it (4) Measure of the EW of the two strongest CaT lines} -
Following the prescriptions presented in \citet{Husser2020}, we
computed the EW of the CaT lines for all the member stars selected in
step (3). First, we normalized the spectra by fitting a second-degree
polynomial to the region of the spectrum that they adopted for the
definition of the continuum (specifically, 8674-8484 \AA, 8563-8577
\AA, 8619-8642 \AA, 8700-8725 \AA, and 8776-8692 \AA). According to
their approach, we used as ``observational quantity'' the sum of the
equivalent widths ($\Sigma {\rm EW}$) of the two broader lines at 8542
\AA\ and 8662 \AA.  Hence, we fit a Voigt profile to each of the two
considered lines in the wavelength ranges 8522-8562 \AA\ and 8642-8682
\AA, respectively. The integration of the best-fit model in the
considered wavelength ranges yields the EW of the two lines, and
finally $\Sigma {\rm EW}$.  To determine the parameters of the best-fit Voigt function,
as well as the uncertainties associated to the measured EWs, we
applied the Markov Chain Monte Carlo (MCMC) sampling technique using
the \texttt{emcee} code \citep{emcee} to sample the posterior
probability distribution. We assumed the following log-likelihood:
\begin{equation}
    \ln\mathcal{L} \propto -\frac{\chi^2}{2} = -\sum_{i=1}^N \Biggl(
    \frac{F_{{\rm mod},i} - F_{{\rm obs},i}} {\sqrt 2 \delta F_{{\rm
          obs},i}} \Biggr)^2,
\label{eq: log-like}    
\end{equation}
where $N$ is the number of pixels in the line bandpass, $F_{{\rm
    mod},i}$ is the i-th value of the Voigt model, $F_{{\rm obs},i}$
is the $i$th value of the normalized observed flux, and $\delta
F_{{\rm obs},i}$ is the associated flux uncertainty (which is provided
in output by PampelMuse) once normalized to the same observed
continuum.

%For each line, we ran 15 chains evolved for 10000 steps,
%adopting flat priors on the Voigt function parameters. We discarded
%the first 3000 steps and we considered the remaining draws for
%building the posterior distributions over the parameters. Randomly
%choosing the parameters from the chain, we built the distribution of
%Voigt functions and we assumed as best-fit Voigt model the one
%corresponding to the 50th percentile, while the corresponding errors
%are the 16th and 84th percentiles.  
For each absorption line, we run the MCMC adopting flat priors on the Voigt function parameters. Every 500 steps of the final chain, we extracted the corresponding function parameters and calculated the relative EW. We assumed as best-fit Voigt model the one corresponding to the 50th percentile, while the errors correspond to the 16th and 84th percentiles.
This analysis suggested that to
minimize the overall uncertainty in the measure of the EWs, a spectrum
with a relatively high $S/N$ ratio is required.  For this reason we
conservatively decided to limit the quantitative measure of the EWs
only to the stellar spectra with $S/N\ge15$.  This assumption sensibly
reduces the number of stars actually measured in each cluster but it
guarantees an appropriate characterization of the continuum and a
solid evaluation of the CaT line intensities.  For the sake of
illustration, in Figure \ref{fig: fit_spectra} we show four normalized
spectra (one for each cluster) with comparable $S/N$ ratios, and the
corresponding best-fit models.

{\it (5) Determination of the star metallicity} - Following
\citet{Husser2020}, we then computed for each star the mean reduced EW
($W'$), which is defined as:
\begin{equation}
    W' = \Sigma \text{EW} + 0.442 (V-V_{\rm HB}) - 0.058 (V-V_{\rm
      HB})^2,
\label{eq: W'}
\end{equation}
where and $V-V_{\rm HB}$ is the difference in magnitude between every
considered star and the cluster HB level, both measured in the F606W
HST filter. The definition of the mean reduced EW therefore includes a
combination of the measured $\Sigma \rm{EW}$ and the brightness of the
observed stars, under the assumption that, at fixed metallicity, the
strength of the CaT lines mainly depends on the stellar luminosity.
To calibrate the relation between $W'$ and [Fe/H], \cite{Husser2020}
adopted the metallicity values listed by \cite{Dias2016} for a sample
of 19 Galactic GCs, and they presented three possible solutions
corresponding to a linear, quadratic, and cubic best-fit to the data.
In the following analysis we adopted the linear calibration:
\begin{equation}
    \text{[Fe/H]} = (-3.61\pm0.13) + (0.52\pm0.03) W'.
\label{eq: FeH}
\end{equation}
This is indeed the most conservative and safe assumption for a work
devoted to the exploration of the metal-rich end of the GC iron
distribution. In fact, the relations of \citet{Husser2020} must be
essentially extrapolated at [Fe/H]$> -0.5$, because this metallicity
regime is not properly sampled by their calibrators. However, it is
well known that extrapolating a quadratic or cubic relation is always
much more dangerous than extrapolating a linear one. In addition, the
iron abundance of their most metal-rich calibrator (NGC 6624) possibly
is significantly overestimated: while \citet{Husser2020} adopted
[Fe/H]$ = -0.36$ from the compilation of \citet{Dias2016},
high-resolution spectra provide [Fe/H]$=-0.69 \pm 0.06$ dex for this
cluster \citep{Valenti6624_6569}.
%% However, we noted that in the metal-rich regime the quadratic and the
%% cubic fits are mainly driven by the metallicity of NGC 6624, which is
%% quoted to be $-0.36\,$dex in the study of \cite{Dias2016}. Since this
%% value is quite in disagreement with high-resolution spectroscopic
%% determinations ($\text{[Fe/H]}=-0.69\pm0.06$ dex;
%% \citealt{Valenti6624_6569}), we decided to adopt the linear
%% calibration:
%% \begin{equation}
%%     \text{[Fe/H]} = (-3.61\pm0.13) + (0.52\pm0.03) W'.
%% \label{eq: FeH}
%% \end{equation}

The error on the individual [Fe/H] measure was estimated computing the quadratic sum of the propagated uncertainty and the root mean square of Equation~\ref{eq: FeH} (see \cite{Husser2020} for details). Considering the assumed $S/N$ ratio cut, the median uncertainty on the individual [Fe/H] measure turns out $\sim0.15\,$dex. 

The magnitude of the HB was estimated as the mean value of the stars
observed along this evolutionary sequence in the differentially
reddening corrected (DRC) CMD. In the case of NGC 6569, for which only
magnitudes in the F555W filter were available, we converted the
measured magnitude difference into the F606W filter by using PARSEC
isochrones \citep{PARSEC} computed considering the cluster parameters
obtained by \cite{Saracino6569}.  We have verified that the derived
metallicity distribution remains unchanged even if the $V-V_{\rm HB}$
magnitude differences are computed in the F555W filter.

\section{Results and discussion }
\label{sec:results}

\subsection{Validation benchmarks}
\label{sec:validations_results} 
As mentioned, before estimating the metallicity
distribution of Liller 1, we tested the validity of the method in a
metallicity regime suitable for bulge star clusters. To this purpose,
we first analyzed the selected benchmark GCs (NGC 6569, NGC 6440 and
NGC 6528), for which spectroscopic values of [Fe/H] are available and
can thus be compared with those obtained from Equation~\ref{eq: FeH}.

As discussed in Section \ref{sec:data_analysis}, we considered as
bona fide targets for the metallicity analysis only the stars with
$S/N\geq15$. Moreover, we excluded saturated objects in the F606W
filter, since Equation~\ref{eq: FeH} requires reliable measures of the
star magnitude in this band.
%In the case of NGC 6569, we also did not consider stars belonging to
%the extended HB because of the blending between the CaT lines and
%%the hydrogen Paschen lines.
This selection let to a final sample of 108 stars for NGC 6569, 100
stars for NGC 6440 and 199 stars for NGC 6528.  The metallicity
distributions derived for each cluster are shown in the left panels of
Figures \ref{fig: NGC6569_MDF}, \ref{fig: NGC6440_MDF} and \ref{fig:
  NGC6528_MDF}, together with the locations of the adopted star
samples in the PM-selected and DRC CMD, which are provided in the
right panels.

As can be seen, the derived metallicity distributions are fully
compatible with a single peak component in all the three reference
clusters. This is also confirmed by the Gaussian mixture model (GMM)
statistics, computed by using the scikit-learn python package
\citep{scikit-learn}.  We led the code free to explore from 1 to 4
components during the fit to the derived metallicity distribution. In
all the cases, both the Bayesian (BIC) and the Akaike (AIC)
information criteria gave as best result a single component. In
addition, for all our benchmark clusters, we obtained a metallicity
determination nicely in agreement with the literature
values. Specifically, for NGC 6569 we obtained $\text{[Fe/H]}= -0.9\,$dex,
with a dispersion $\sigma = 0.24$. This value is in good agreement
with high-resolution spectroscopic determinations, which range from
$\text{[Fe/H]} = -0.87\,$dex \citep{Johnson6569} to $\text{[Fe/H]} = -0.79\,$dex
\citep{Valenti6624_6569}.  In the case of NGC 6440, we find that the
mean of the distribution is equal to $\text{[Fe/H]} = -0.53\,$dex, with
$\sigma = 0.2$, which in excellent agreement with high-resolution
spectroscopy studies: $\text{[Fe/H]} = -0.5\,$dex \citep{Munoz6440}, $\text{[Fe/H]} =
-0.56\,$dex \citep{Origlia6440}.  Finally, for NGC 6528 we find a mean
value of $\text{[Fe/H]} = -0.23\,$dex with $\sigma = 0.19$. Taking into
account the (large) dispersion of the distribution, we conclude that
also in this case the derived value is in satisfactory agreement with
previous measures from high-resolution spectra: \cite{Munoz6528} and
\cite{Origlia6528} quote $\text{[Fe/H]} = -0.2\,$dex and $\text{[Fe/H]} = -0.17\,$dex,
respectively, and \cite{Schiavon6528} obtained a value of $\text{[Fe/H]} \sim
-0.2\,$dex from two APOGEE \citep{APOGEE} stars.
%Given that the typical uncertainty on the individual [Fe/H] estimates is $\sim0.15\,$dex, we conclude
% that the width of the iron distributions is not due to an intrinsic iron spread of the stellar populations.

Thus, the first results of this analysis is that the approach proposed
by \cite{Husser2020} to determine the metal content of star clusters
from the CaT lines, and the linear calibration shown in Equation~\ref{eq: FeH} hold also in the high-metallicity regime typical of bulge GCs,
yielding iron distributions fully in agreement with the literature.

\begin{figure}[t]
    \centering
    \includegraphics[width=0.8\textwidth]{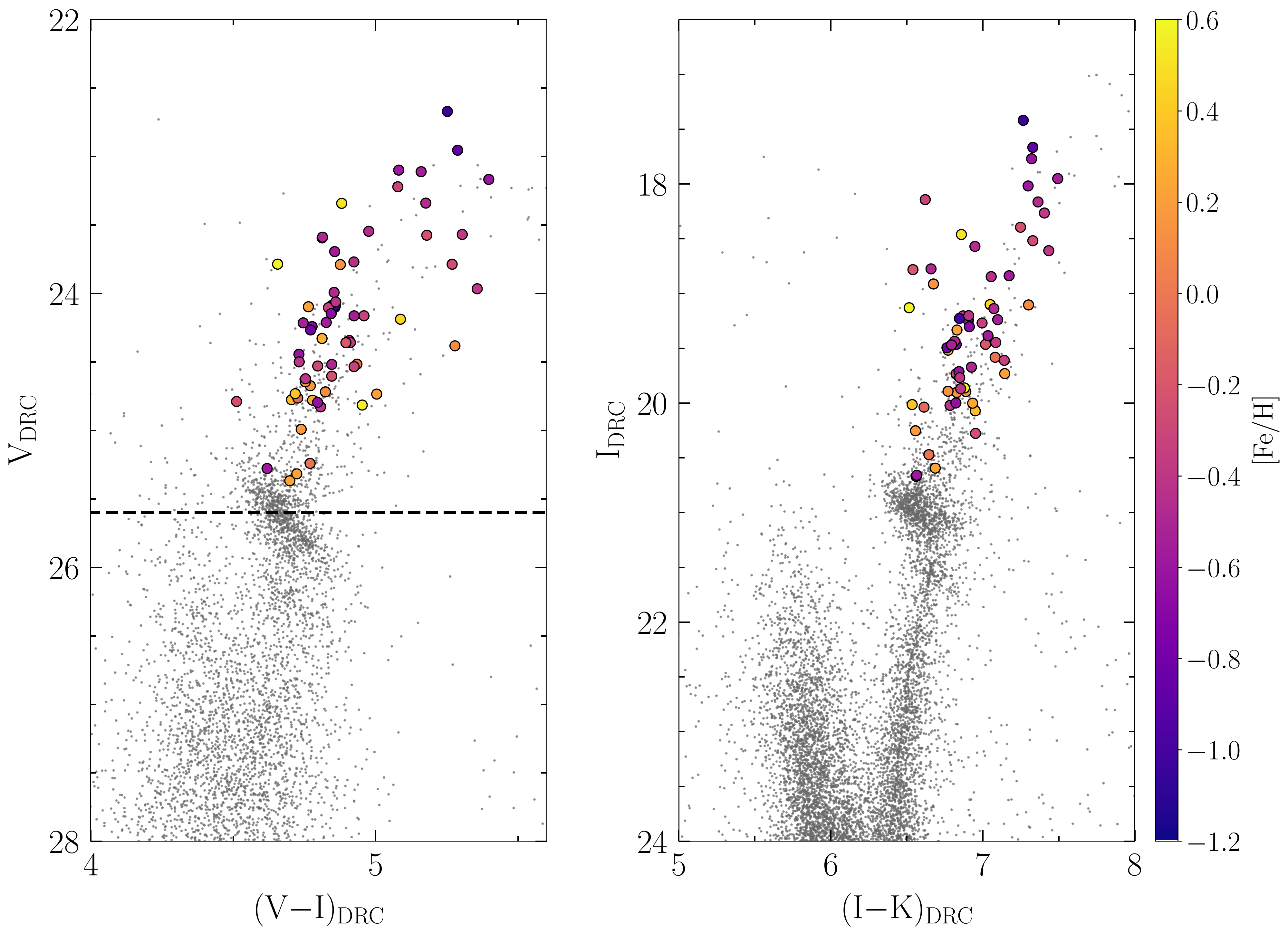}
    \caption{PM-selected and DRC optical and hybrid CMDs of Liller 1
      (black dots) from the photometric catalog presented in
      \cite{ferraro+21}.  The stars with measured metallicity are
      plotted as circles colored according to their [Fe/H] (see the
      color bar on the right). The black dashed line in the left panel
      marks the adopted HB level optical magnitude ($V_{\rm HB} = 25.6$).}
    \label{fig: Liller1_CMDs}
\end{figure}

\begin{figure}[t]
    \centering
    \includegraphics[width=0.6\textwidth]{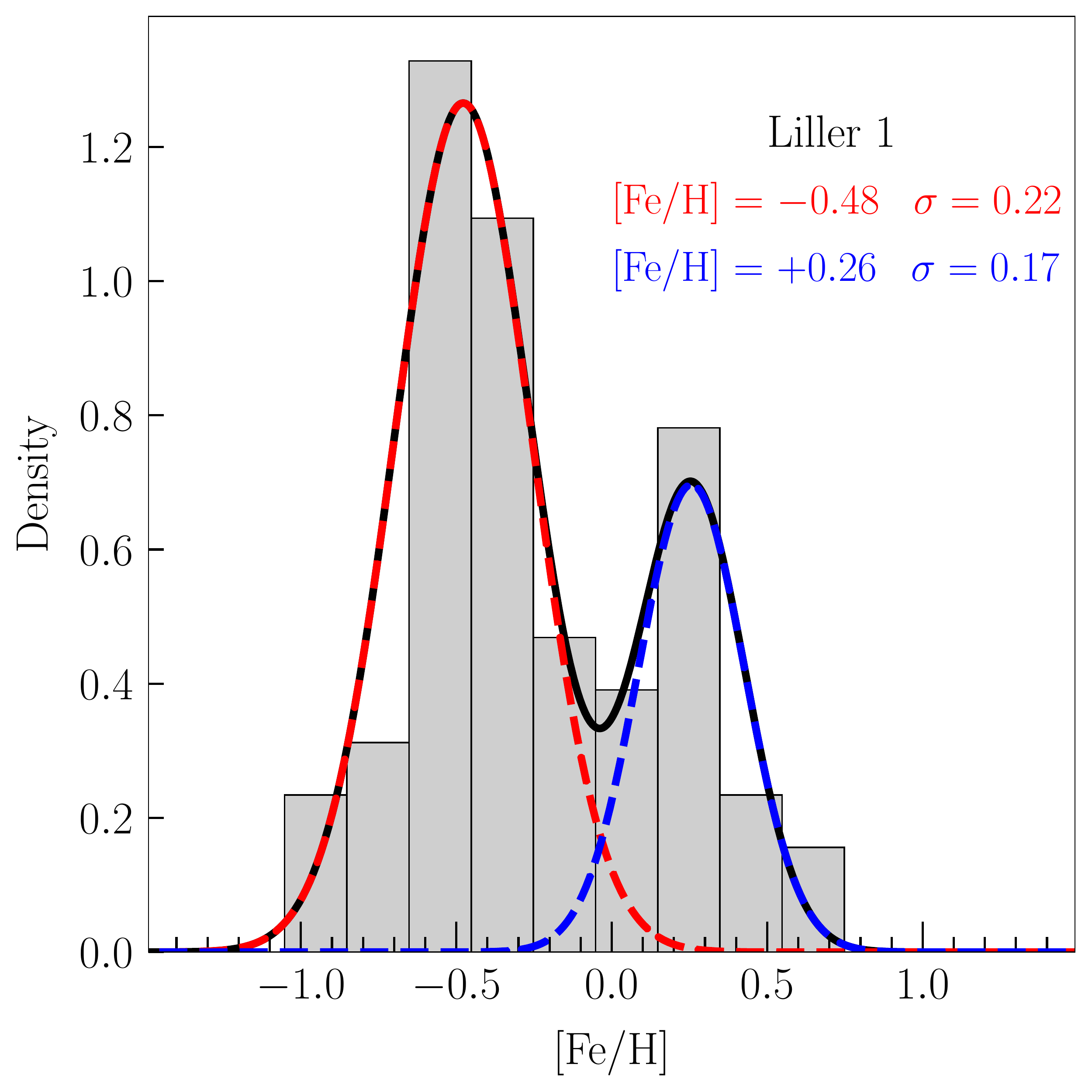}
    \caption{Metallicity distribution of Liller 1 bona-fide targets (grey histogram).
      The solid black line shows the function that best reproduces the
      observed distribution. It is the combination of the two Gaussian
      functions shown as red and blue dashed lines and indicating the
      presence, respectively, of a metal-poor and a metal-rich
      sub-populations in Liller 1.  The mean [Fe/H] values and the
      standard deviations of the two individual Gaussian components
      are also labelled in the panel. }
    \label{fig: Liller1_MDF}
\end{figure}

\subsection{Liller 1}
\label{sec:liller1_results}
To determine the metallicity distribution of Liller 1, we selected the
stars according to the criteria described in Section
\ref{sec:validations_results} and we further excluded spectra of
extremely cool objects, with colors $(I-K)\geq7.5$ (see Figure
\ref{fig: Liller1_CMDs}), because strong TiO molecular bands fall in
the CaT wavelenght range.  After this selection, the final sample
counts 64 stars.  The positions of these objects in the PM-selected
and DRC optical ($V, V-I$) and hybrid ($I, I-K$) CMDs are shown in
Figure \ref{fig: Liller1_CMDs}, where the points are color coded on the
basis of their derived metallicity. As apparent, at odds with what
found for the benchmark GCs, a significant iron spread is already
appreciable from this plot.

Figure \ref{fig: Liller1_MDF} shows the obtained metallicity
distribution. In spite of the same $S/N$ cut and the same bin size,
the histogram of Liller 1 is clearly different from those presented in
Section \ref{sec:validations_results} for the three reference
clusters, showing a clear bimodality, with at least 14 stars having
super-solar metallicity.  We checked for possible spurious effects
that could have artificially generated the metal-rich peak observed in
Figure \ref{fig: Liller1_MDF}.
%First, we verified the absence of trends with the brightness
%difference to the HB, i.e. with $T_{eff}$. This is shown in
%%Figure~\ref{fig: Liller1_VHB_FeH}, in which we plot the [Fe/H]
%values versus the difference $V-V_{HB}$, where we do not appreciate
%%any trend.  Second,
First, we noted that the vast majority (12 out of 14) of the
identified super-solar stars have both RV and PM measurements, thus
guaranteeing their membership to Liller 1. This is also supported by
their radial distribution, discussed in the next section.  We also
verified the absence of trends with the differential reddening value
used to correct the F606W magnitudes that enter Equation~\ref{eq: W'}. In
fact, Liller 1 is one of the most extincted bulge stellar clusters,
with an average $E(B-V) = 4.52\pm0.10$ and a maximum variation of
reddening of $\delta E(B-V)\sim 0.9$ due to highly spatially variable
interstellar extinction in its direction
\citep{Pallanca_Liller1_redd}. This phenomenon heavily affects
magnitudes, especially in the optical bands. Figure \ref{fig:
  Liller1_debv_FeH} demonstrates that there is no trend between
$\delta E(B-V)$ and the derived metallicity values. 
This is also confirmed by the Pearson coefficient which turns out to be $\sim-0.2$,
thus confirming the absence of correlations among the two quantities.
Thus, we can
safely exclude that the metal rich component is due to any bias in the
measures.

To statistically verify the multi-modality of the iron distribution of
Liller 1, we run the GMM code letting it free to assess the number of
components. Indeed, both the BIC and the AIC analysis give the best
result for two Gaussian components.  Considering the mean values and
dispersions obtained from the GMM, the two peaks correspond to
$\text{[Fe/H]} = -0.48\,$dex (with $\sigma = 0.22$) and $\text{[Fe/H]}
= +0.26\,$dex (with $\sigma = 0.17$).  The two components that
reproduce the overall shape of the iron distribution of Liller 1 are
shown with different colors in Figure \ref{fig: Liller1_MDF}.  This is
the first spectroscopic confirmation of a super-solar stellar
population in Liller 1, as strongly suggested by the photometric
analysis presented in \cite{ferraro+21} and \citet{dalessandro+22}.
We emphasize that a bimodal distribution is found also if the
  quadratic calibration of \citet{Husser2020} is adopted, in place of
  the linear one. The only variations are that both the metal-rich and
  the metal-poor components show larger dispersions, and the
  metal-rich peak is found at [Fe/H]$\simeq 1$ (indeed, such an
  unrealistically high value of iron abundance further suggests that
  the linear calibration is the most appropriate choice in the
  high-metallicity regime; see the discussion in Section
  \ref{sec:data_analysis}).

\begin{figure}[t]
    \centering
    \includegraphics[width=0.6\textwidth]{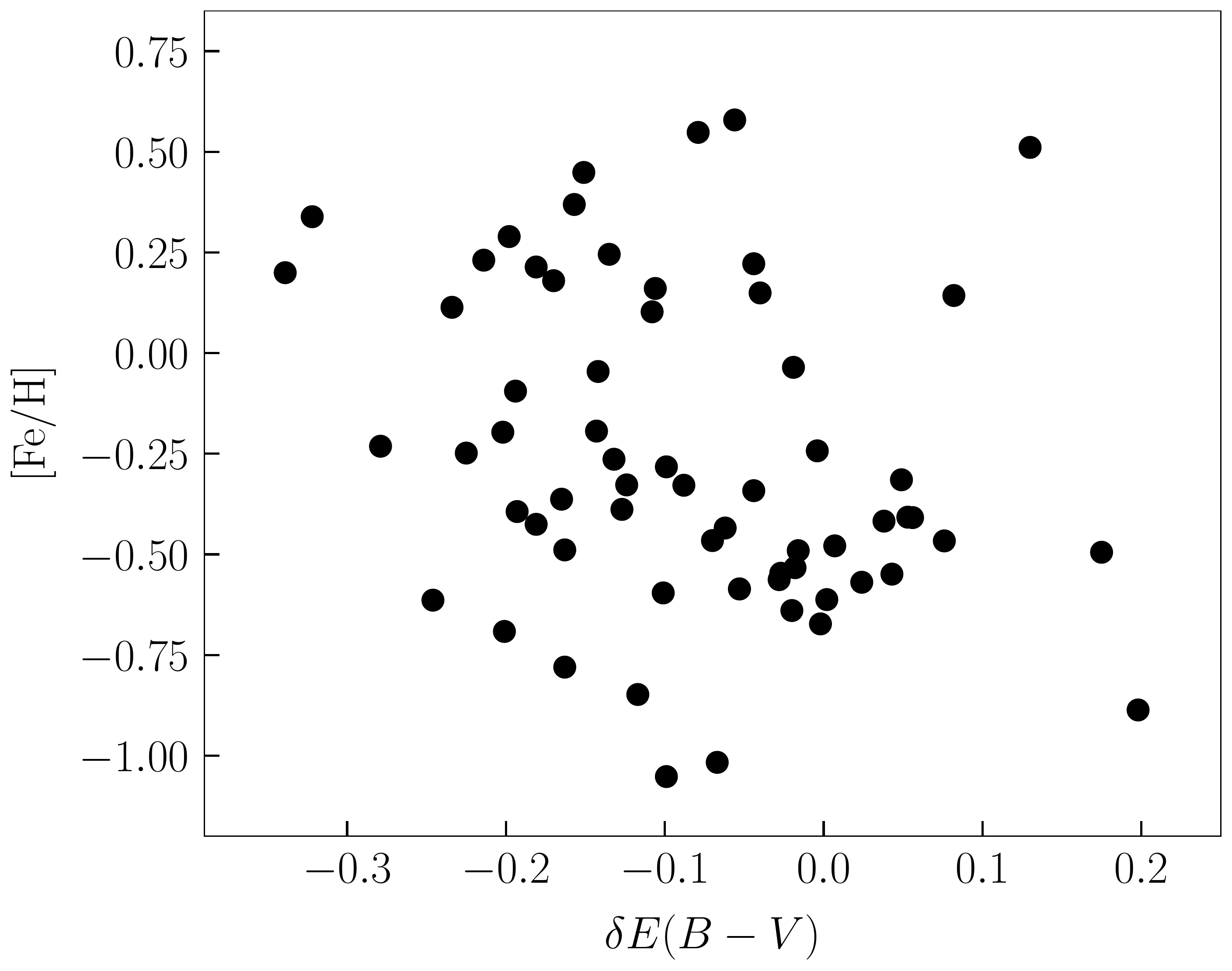}
    \caption{Measured values of [Fe/H] as a function the differential
      color excess $\delta E(B-V)$ (from
      \citealt{Pallanca_Liller1_redd}) used to build the DRC CMD of
      Liller 1.}
    \label{fig: Liller1_debv_FeH}
\end{figure}

\section{Summary and conclusions }
\label{sec:summary_conclusions}
In this study we present the first spectroscopic metallicity distribution ever
obtained for the bulge stellar system Liller 1. This has been
determined from the analysis of MUSE spectra of 64 individual member
stars, and the measure of the EW of the CaT lines following the
prescriptions described in \cite{Husser2020}.

These authors presented three relations linking the CaT EW to [Fe/H],
obtained from linear, quadratic, and cubic fits to the data provided
by their calibration sample (19 Galactic GCs from
\citealp{Dias2016}). However, since the metal rich regime is poorly
represented in that sample\footnote{Specifically, four GCs are considered in the metallicity regime of interest for this study, namely NGC 104, NGC 6388, NGC 6441 and NGC 6624. Excluding NGC 6624, for which high-resolution spectroscopic measurements disagree with the metallicity value derived in \cite{Dias2016}, the metal-rich sample considered by \cite{Husser2020} spans a metallicity range from $\sim-0.7\,$dex up to $\sim-0.5\,$dex.}, to avoid hazardous extrapolations we have
adopted their linear calibration (see Equation~\ref{eq: FeH}). Then, to test its
validity in the high metallicity regime, as a sanity check, we first
determined the iron distribution of three reference bulge GCs of known
metallicity (namely NGC 6569, NGC 6440, and NGC 6528).  In all cases
we found unimodal distributions peaked at a mean iron abundance that
is fully consistent with the spectroscopic values reported in the
literature. 
%\textbf{Moreover, the dispersion of the distributions can be ascribed to the typical uncertainty on the abundance %estimates.}
%% : NGC 6569 ($\text{[Fe/H]} = -0.9\,\text{dex}$, $\sigma = 0.24$), NGC
%% 6440 ($\text{[Fe/H]} = -0.53\,\text{dex}$, $\sigma = 0.2$) and NGC
%% 6528 ($\text{[Fe/H]} = -0.23\,\text{dex}$, $\sigma = 0.19$).

The application of the same methodology to Liller 1 yielded, instead,
a completely different result: a clear bimodal iron distribution is
obtained, with a main peak at $\text{[Fe/H]}\sim -0.5\,$dex and a secondary peak
at $\text{[Fe/H]}\sim +0.3\,$dex. Indeed, both the BIC and the AIC analyses
confirm that the overall distribution is best represented by a
combination of two Gaussian components.
The metallicity of the metal-poor population is fully consistent with
the value quoted by \citet{origlia+02} ([Fe/H] $= -0.3\pm0.2\,$dex),
while the secondary peak detected here is the first spectroscopic
confirmation of the presence of a super-solar stellar population in
Liller 1. This finding is in perfect agreement with what suggested by
the photometric analyses presented in \cite{ferraro+21} and \cite{dalessandro+22}.

\begin{figure}[t]
\begin{minipage}{0.5\linewidth}
  %\rule{\linewidth}{0.75\linewidth}
   \includegraphics[width=0.9\textwidth]{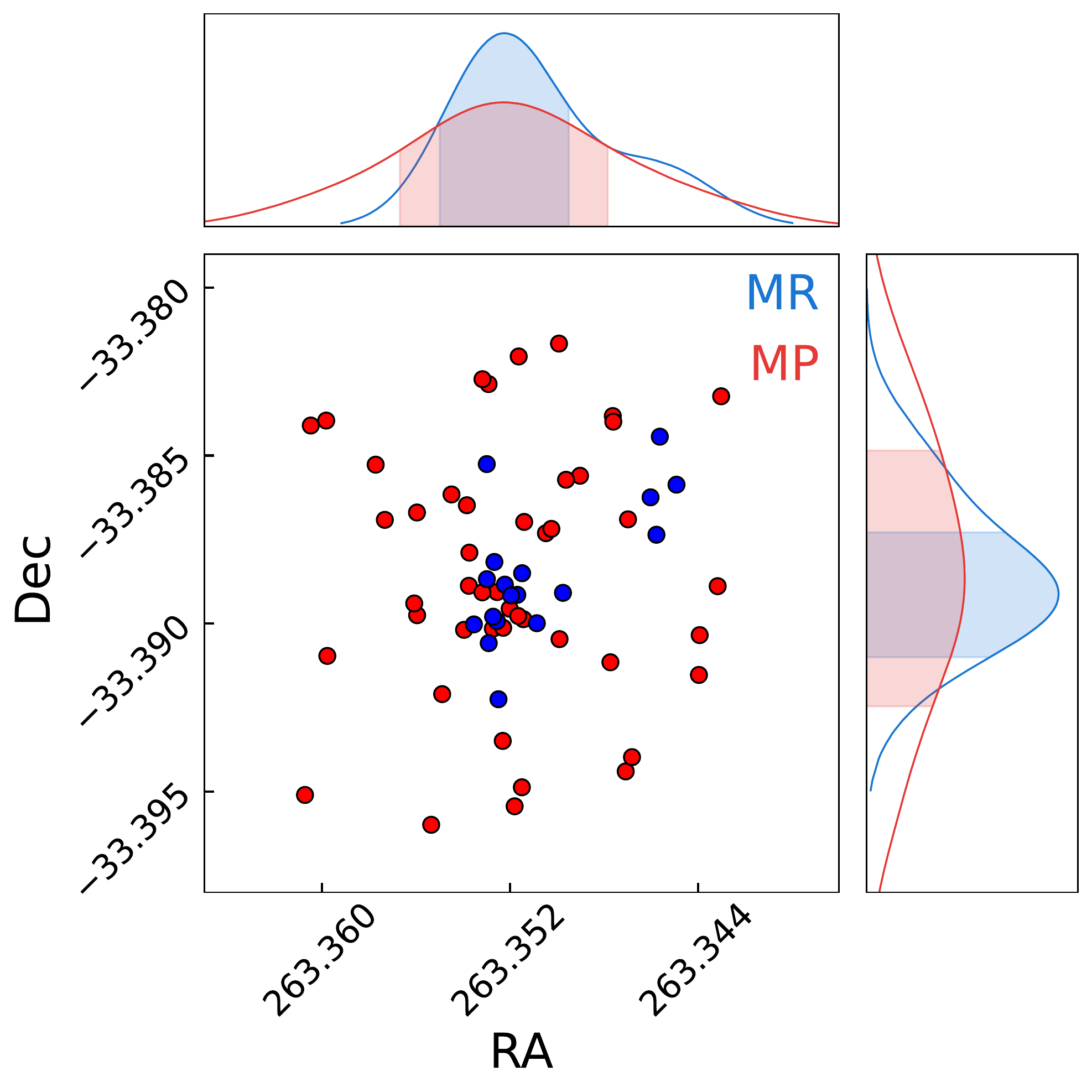}
  %\caption{CMD of Terzan 5 from Ferraro+16.}\label{fig:dummy-1}
\end{minipage}
\hfill
\begin{minipage}{0.45\linewidth}
  %\rule{\linewidth}{0.75\linewidth}
    \includegraphics[width=0.88\textwidth]{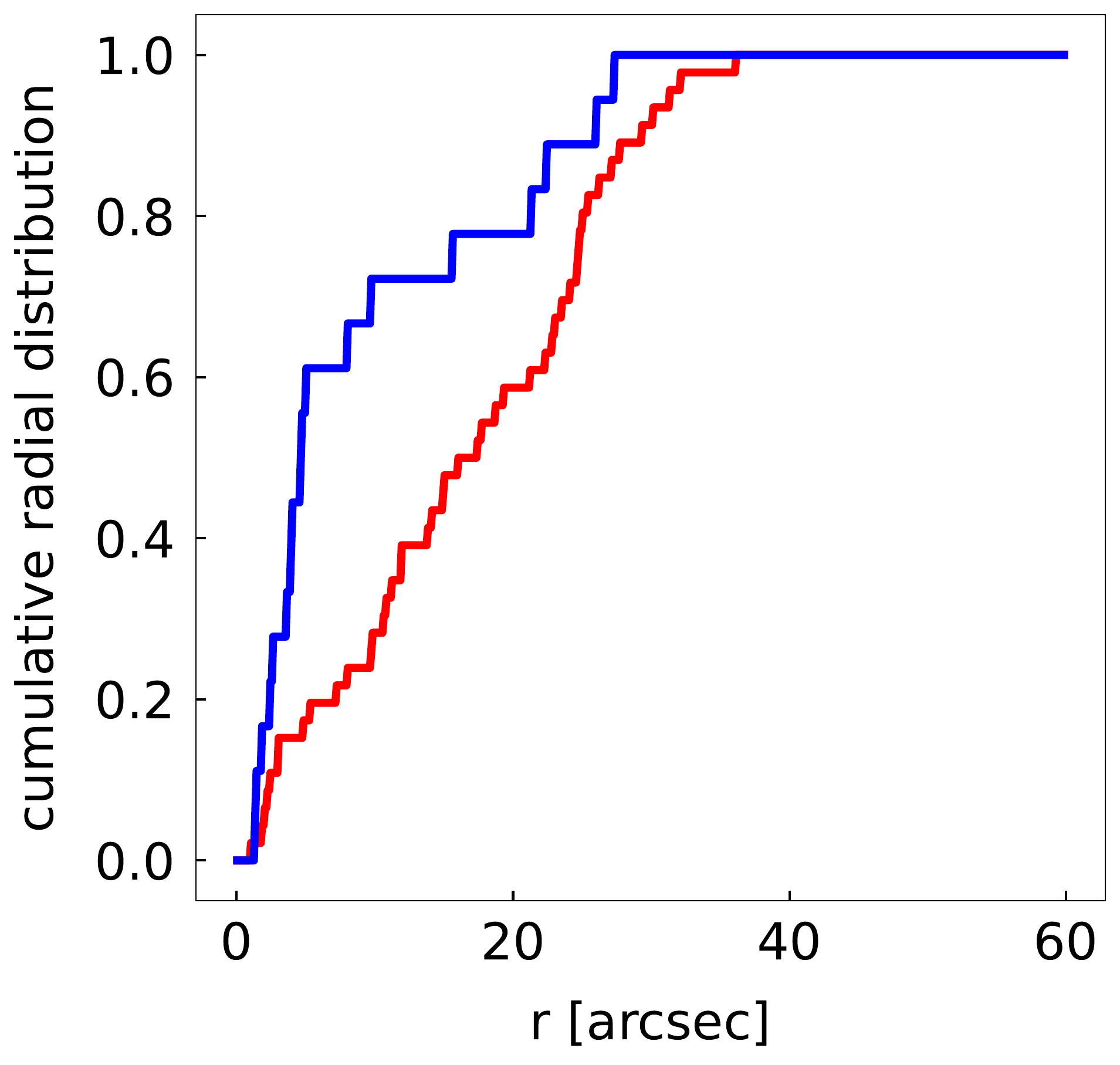}
  %\caption{CM}\label{fig:dummy-2}
\end{minipage}
\caption{\emph{Left panel:} Spatial distribution of the stars with
  measured metallicity in Liller 1.  The red circles mark the stars
  with a probability $\geq 0.5$ to belong to the metal-poor (MP)
  component, the blue circles are those with a probability $\geq 0.5$ to
  belong to the metal-rich (MR) sub-population.  The projected 1D
  distributions of the two sub-samples, along the right ascension and
  the declination directions, are shown in the top and right side
  panels, respectively, with the shaded areas corresponding to the
  $1\sigma$ confidence level of the distributions. \emph{Righ panel:}
  Cumulative radial distributions of the MP (red line) and the MR
  (blue line) sub-populations of Liller 1.}
\label{fig: spatial_distr}
\end{figure}

The modelling of the observed metallicity distribution with the two
Gaussian components discussed above (Section
\ref{sec:liller1_results}) allows us to assign to each star a
probability of belonging to the sub-solar or to the super-solar
component. With this additional information we can probe the spatial
distribution of the stars belonging to the two populations. Very
interestingly, we find that stars with a probability larger than 0.5
to belong to the super-solar component appear more centrally
concentrated than the sub-solar population. This is shown in Figure
\ref{fig: spatial_distr}, where the difference between the two
distribution is clearly distinguishable also by eyes.  The cumulative
radial distributions of the two components fully confirm this
indication, and allow us to evaluate the statistical significance of
the difference. The Kolmogorov-Smirnov test applied to the two
components shows that the probability that they are extracted from the
same parent distribution is essentially zero (p=0.000), indicating that
the two distributions are different at more than $5\sigma$ of
statistical significance. This finding is strikingly similar to what
found in another bulge stellar system, Terzan 5 \citep{ferraro+09,
  lanzoni+10}. Indeed, the two main components of the metallicity
distribution of Terzan 5 share exactly the same behavior, with the
metal-rich population being more centrally segregated than the metal
poor one. This strongly reinforces the similarity between Terzan 5 and
Liller 1 and, in both cases, it is strongly evocative of a
self-enrichment scenario, where the more-metal rich component formed
from gas ejected by SN explosions, that was retained by the stellar
system and progressively deposited in its central region.

In this respect, the star formation history of Liller 1, recently
reconstructed by \cite{dalessandro+22} from the CMD position of the
member stars selected in \cite{ferraro+21}, suggests that it was
characterized by three main bursts (occurred 12-13 Gyr ago, 6-9 Gyr
ago and, the most recent one, only 1-3 Gyr ago) combined with a low,
but constant, activity of star formation over the entire lifetime of
the system. The predicted metallicity distribution shows two main
peaks (at $\text{[M/H]}=-0.5\,$dex and $\text{[M/H]}=+0.2\,$dex) and is astonishingly similar
to that derived here (compare Figure \ref{fig: Liller1_MDF} with Figure
8 in \citealp{dalessandro+22}).  Thus, while the results discussed in
\cite{dalessandro+22} demonstrate that Liller 1 unlikely formed
through the merger between an old GC and a giant molecular cloud (as
it was recently proposed in \citealp{bastian+22}), the findings
presented in this paper provide the first spectroscopic evidence of
the presence of a super-solar component in this system, adding further
support to the idea that Liller 1 is the surviving relic of a massive
primordial structure (similar to the giant clumps observed in
star-forming high-redshift galaxies) that contributed to the formation
of the Galactic bulge.

The sample of stars measured in this study provides the primary target
list for a detailed high-resolution spectroscopic screening of key
chemical elements (iron-peak, $\alpha$-elements, etc.) that are needed
for the full reconstruction of the enrichment history of Liller 1.
Indeed, the approach adopted in this work represents a methodological
reference benchmark for the study of any additional Bulge Fossil
Fragment that will be discovered in the future: it provides
an efficient way to determine a preliminary (but meaningful)
metallicity distribution, which is able to unveil the presence of
multi-iron sub-populations and to provide a list of top-priority
targets for follow-up high-resolution spectroscopic investigations. On
the other hand, the combined spectroscopic and photometric search for
new Bulge Fossil Fragments is of primary importance to clarify the
formation mechanism of our Galaxy. In fact, solid assessing the amount
of stellar systems belonging to this new class of objects can provide
invaluable pieces of information about the role played by merging
processes in the formation of the Milky Way bulge.

\vskip1truecm
CC ackwnoledges the Marco Polo grant for funding the period she spent
at the ESO Headquartier in Garching bei M\"{u}nchen (Germany) where part
of this work was carried out. CC also kindly thanks Dr.~S. Kamann for useful suggestions about PampelMuse. This research is part of the project
{\it Cosmic-Lab} (“Globular Clusters as Cosmic Laboratories”)
at the Physics and Astronomy Department “A. Righi” of
the Bologna University (http://www.cosmic-lab.eu/
Cosmic-Lab/Home.html). The research was funded by the MIUR throughout
the PRIN-2017 grant awarded to the project {\it Light-on-Dark}
(PI:Ferraro) through contract PRIN-2017K7REXT.
EV acknowledges the Excellence Cluster ORIGINS Funded by the Deutsche Forschungsgemeinschaft (DFG, German Research Foundation) under Germany’s Excellence Strategy – EXC-2094 – 390783311.

\newpage
%\bibliography{bibliografia}{}
%\bibliographystyle{aasjournal}

\end{document}